\documentclass[12pt,letterpaper,hyper]{JHEP3}
\usepackage{amsmath,epsfig}
\usepackage{amssymb,amsfonts}
\usepackage{graphicx}
\usepackage{multirow,cite}
\usepackage{subfigure}

\title{No entropy enigmas for   $\N=4$ dyons }
\preprint{}
\author{
Atish Dabholkar$^{1, 2}$, Monica Guica$^{1}$, Sameer Murthy$^{1}$ and Suresh Nampuri$^{2,3}$\\

\it $^1${Laboratoire de Physique Th\'eorique et Hautes Energies (LPTHE)\\
\it{Universit\'e Pierre et Marie Curie-Paris 6; CNRS UMR 7589}\\
\it{Tour 24-25, 5$^{\grave{e}me}$ \'etage, Boite 126, 4 Place Jussieu} \\
\it {75252 Paris Cedex 05, France}}\\

\it $^2$Department of Theoretical Physics\\
\it Tata Institute of Fundamental Research\\
\it Homi Bhabha Rd, Mumbai 400 005, India\\

\it $^3$Arnold Sommerfeld Centre for Theoretical Physics\\
\it Ludwig-Maximilians-Universit\"{a}t M\"{u}nchen\\
\it Department f\"{u}r Physik\\
\it Theresienstr. 37, 80333 M\"{u}nchen, Germany\\

}

\abstract{We explain why multi-centered black hole configurations where at least one of the centers is a large black hole do not contribute to the indexed degeneracies in theories with $\CN=4$ supersymmetry. This is a consequence of the fact that such configurations, although  supersymmetric, belong to long supermultiplets. As a result, there is no entropy enigma in $\CN=4$ theories, unlike in $\CN=2$ theories.}

\keywords{black holes, superstrings, dyons}

\newenvironment{myenumerate}{
\begin{enumerate}
   \setlength{\itemsep}{1pt}
   \setlength{\parskip}{0pt}
   \setlength{\parsep}{0pt}}{\end{enumerate}}

\newcommand{\bi}{\begin{itemize}}
\newcommand{\ei}{\end{itemize}}

\newcommand{\IR}{\mathbb{R}}

\def\d{\partial}

\def\a{\alpha}
\def\b{\beta}
\def\e{\epsilon}

\def\h{\eta}

\def\CM{{\cal M}}

\def\CN{{\cal N}}

\def\half{{\frac12}}

\def\CN{{\cal N}}

\def\1F1{{}_1\!F_1}
\def\2F0{{}_2\!F_0}

\def\a{\alpha}

\def\h3{$\textrm{H}_3^+$}

\def\d{{\partial}}

\def\IR{{\mathbb R}}

\def\CM{{\cal M}}
\def\CN{{\cal N}}

\font\manual=manfnt
\def\dbend{\lower3.5pt\hbox{\manual\char127}}

\def\bar{\overline}

\def\CN{{\cal N}}

\def\rt2{\sqrt{2}}
\def\irt2{{1\over\sqrt{2}}}

\def\b{\beta}
\def\a{\alpha}

\font\cmss=cmss10
\font\cmsss=cmss10 at 7pt
\def\IL{\relax{\rm I\kern-.18em L}}
\def\IH{\relax{\rm I\kern-.18em H}}
\def\rlx{\relax\leavevmode}
\def\ZZ{\rlx\leavevmode\ifmmode\mathchoice{\hbox{\cmss Z\kern-.4em Z}}
 {\hbox{\cmss Z\kern-.4em Z}}{\lower.9pt\hbox{\cmsss Z\kern-.36em Z}}
 {\lower1.2pt\hbox{\cmsss Z\kern-.36em Z}}\else{\cmss Z\kern-.4em
 Z}\fi}

\def\rt2{\sqrt{2}}
\def\irt2{{1\over\sqrt{2}}}

\newcommand{\Z}{{\mathbb Z}}
\newcommand{\R}{{\mathbb R}}
\newcommand{\N}{{\mathcal{N}}}
\newcommand{\C}{{\mathbb C}}

\newcommand{\bea}{\begin{eqnarray}}
\newcommand{\eea}{\end{eqnarray}}
\newcommand{\be}{\begin{equation}}
\newcommand{\ee}{\end{equation}}
\newcommand{\ben}{\begin{eqnarray*}}
\newcommand{\een}{\end{eqnarray*}}
\newcommand{\bem}{\begin{pmatrix}}
\newcommand{\eem}{\end{pmatrix}}
\newcommand{\bl}{\begin{align}}
\newcommand{\el}{\end{align}}

\def\a{\alpha}
\def\b{\beta}

\def\d{\delta}

\def\e{\epsilon}

\def\h{\eta}

\begin{document}


\subsection*{An Enigma}

To obtain a microscopic quantum description of supersymmetric black holes in string theory, one usually starts at weak coupling with a brane configuration of given mass and charges localized at a single point in the noncompact spacetime. One then computes an appropriate indexed partition function in the world-volume theory of the branes. In spacetime, this index corresponds to the helicity supertrace that counts  BPS supermultiplets. At strong coupling, the brane configuration gravitates and the indexed partition function is expected to count the microstates of these macroscopic gravitating configurations. Assuming that the gravitating configuration is a single-centered black hole, these considerations provide a statistical understanding of the entropy of the black hole in terms of its microstates, in agreement with the Boltzmann relation\footnote{It is usually assumed that the index equals the absolute number, following the dictum that whatever can get paired up will get paired up. This assumption is borne out in several examples but may fail in general.}.

One problem with this approach  is that the macroscopic supergravity solutions corresponding to the microscopic brane configuration need not be centered at a point. Instead, they may include several multi-centered black holes in addition to the single-centered black hole of interest. In fact, in four-dimensional  $\mathcal{N}=2$ supergravity, it is known that in certain situations there are multi-centered configurations which have \textit{more} entropy than the single-centered black holes carrying the same total charges.   This raises the question  as to why the degeneracy extracted from the microscopic counting function should agree with the entropy of just the single-centered black hole, as is the case in many examples. This puzzle has been referred to as the `entropy enigma' \cite{Denef:2007vg,Denef:2007yn}. One expects that the enigmatic multi-centered solutions in $\CN=2$ supergravity mentioned above can be embedded in  $\CN=4$ supergravity, and should dominate the entropy of single-centered black holes.  We thus have an $\CN=4$ version of the entropy enigma, which is what we address in this note.

To  formulate the puzzle more precisely in this context, note that in $\CN=4$ supersymmetric theories, a BPS-state may preserve either one-half or one-quarter of the sixteen supercharges. A half-BPS state  generically belongs to  a 16-dimensional short multiplet, whereas a quarter-BPS state belongs to  a 64-dimensional intermediate multiplet.
In several string compactifications with $\CN=4$ supersymmetry, the indexed degeneracies that count the intermediate multiplets are known exactly \cite{Dijkgraaf:1996it, Gaiotto:2005gf, Shih:2005uc, David:2006ud,David:2006ru,David:2006ji,Jatkar:2005bh,
Sen:2007qy,Dabholkar:2008zy,Dabholkar:2008tm,Dabholkar:2007zz,Dabholkar:2007vk,Dabholkar:2006xa,Banerjee:2008pu}. In the limit of large charges, when all charges scale as $\lambda$, the
logarithm of the degeneracy is found to be in precise agreement with the entropy of a single black hole, which scales as $\lambda^2$  \cite{Dijkgraaf:1996it, Cardoso:2006bg, Cardoso:2004xf,deWit:2007dn,David:2006yn}.
The only multi-centered configurations that seem to contribute to the exact formula are bound states of two centers such that each center is individually half-BPS, but together they preserve only a quarter of the supersymmetries.  A half-BPS state necessarily corresponds to a small black hole with a string scale horizon, whose entropy always scales as $\lambda$.
The entropy of the multi-centered configuration then also scales as $\lambda$,  which is small compared to the leading term.

The enigmatic configurations in $\CN=2$ theories, on the other hand, are  multi-centered configurations, where each center is a large black hole.
In $\CN=4$ supergravity, this means that  they must correspond to configurations where at least one of the centers is quarter-BPS.
The enigma  can then be rephrased as the following question\footnote{This is actually a stronger question than the original enigma, which  was why certain large multi-centered configurations not {\it dominate} the degeneracy.
We will actually answer the stronger question here.}: why do the multi-centered configurations with at least one quarter-BPS center not contribute to the index that counts
the intermediate multiplets? In our following discussion in the $\CN=4$ context,  we will loosely refer to any multi-centered configuration as `enigmatic'  if at least one of the centers is a large quarter-BPS black hole.

The explanation of this puzzle is rather simple.  It relies on the fact that even though the enigmatic multi-centered configurations are supersymmetric, they belong to {\em long} multiplets, which are 256-dimensional. This happens because there are additional fermionic zero modes apart from the ones that arise due to supersymmetry breaking. As a result, these configurations give a vanishing contribution to the indexed degeneracies.

While this explanation is not entirely unexpected, it is not easy to  directly ascertain the existence of additional fermionic collective coordinates. To do so, one must solve the Dirac equation and  the Rarita-Schwinger equation for the dilatini and the  gravitini in the background of the multi-centered configuration under consideration. One must then show that these additional fermionic collective coordinates are  free and have a quadratic effective action. It is then possible to show that  the quantization of these additional zero modes will make the index vanish. Although in principle it is possible to follow such a path, in practice it is difficult to execute it for the multi-centered black hole solutions in supergravity.

We will  instead give an indirect argument for the fact that  enigmatic  configurations belong to long multiplets, by  showing  that they are continuously connected to nonsupersymmetric configurations. Nonsupersymmetric solutions obviously belong to long multiplets. For them, it is easy to explicitly establish  the existence of fermionic zero modes -- they arise as goldstini of broken supertranslations.  Continuity then implies that the enigmatic configurations must also belong to long multiplets. As we will see, this fact is consistent with the known pattern of wall-crossings for $\CN=4$ dyons and the pole structure of the dyon partition function.

The basic argument is suggested by another closely related puzzle.
Consider for concreteness the simplest $\CN=4$ compactification of heterotic string theory on $T^6$.  For this compactification, the exact partition function counting $64$-dimensional supermultiplets of quarter-BPS dyons is known for {\it all} dyons in  all duality orbits in all regions of moduli space. At certain points in the string moduli space, the gauge symmetry is enhanced to a  nonabelian group, for example $SU(N)$. Away from these points, the gauge symmetry is broken to $U(1)^N$. If the symmetry breaking mass  scale is much smaller than the string scale, then one can decouple gravity and string modes and analyze the BPS states in the field theory limit.
Quarter-BPS dyons in $SU(N)$ super-Yang-Mills theory are well-studied \cite{Lee:1998nv,Bak:1999ip,Bak:1999vd}, and their exact degeneracies are known \cite{Stern:2000ie, Dabholkar:2008tm}.
One would expect that the exact dyon partition function derived in string theory should correctly reproduce the degeneracies of these field theory dyons. This is indeed the case if the gauge group is $SU(3)$\cite{Sen:2007ri, Dabholkar:2008zy}. However, the string theory counting fails to agree if the gauge group is $SU(N)$ with $N\geq 4$.  The puzzle is then why the string theory partition function does not count the general $SU(N)$ dyons.

The resolution of this puzzle is easier to see in the field theory limit \cite{Bergman:1998gs}. It follows  from the fact that the quarter-BPS dyons in $SU(N)$ super Yang-Mills theory with $\N=4$ supersymmetry and $N \geq 4$ exist as supersymmetric configurations only on a submanifold in  moduli space. Even slightly away from this submanifold, the dyons are no longer supersymmetric and hence belong to long multiplets. One can thus argue by continuity that even on the submanifold where they are supersymmetric,  they must belong to long multiplets. In field theory, it is easy to independently verify this argument by explicitly demonstrating the existence of the required additional fermionic zero modes\footnote{Since our argument relies on special properties of  $\CN=4$, it does not address the original entropy enigma in $\CN=2$ theories. In $\CN=2$ theories, both large and small black holes are half-BPS and one cannot distinguish them by the size of the supermultiplet.}.

In string theory, single centered black hole solutions of the $\N=2$ theory have been embedded into the $\CN=4$ theory using a consistent truncation to the former theory  \cite{LopesCardoso:1999ur}.
One may expect that a similar uplift can be done for the $\CN=2$ multi-centered solutions.  In principle, there could also exist more general solutions which cannot be truncated to the $\CN=2$ theory.
Our arguments apply to the most general case, and show that supersymmetric multi-centered solutions can only exist on submanifolds of  the $\CN=4$ moduli space of codimension greater than one.

In the following, we first review the simpler field theory argument and  then generalize it to  the supergravity case.

\subsection*{Dyons in Field Theory}

The quarter-BPS dyons in  $SU(N)$ $\N=4$ Yang-Mills   have a particularly simple and geometric representation in terms of string webs ending on $N$ D3-branes\cite{Dasgupta:1997pu,Sen:1997xi,Bergman:1997yw,Bergman:1998gs,Schwarz:1996bh,Argyres:2001pv,Argyres:2000xs,Narayan:2007tx}.
There is an overall $U(1)$ of the center-of-mass motion that does not play any role. The transverse space to the D3-branes is $\mathbb{R}^6$, and therefore the moduli space of the Coulomb branch of the $N$ D3 branes is $(\mathbb{R}^6)^N$.
At a generic point in moduli space where all D3-branes are separated from each other, the gauge group is completely broken to $U(1)^N$. The quarter-BPS states in question correspond to a planar three-pronged string junction network stretched between the $N$ D3-branes. Only  planar configurations are supersymmetric \cite{Dasgupta:1997pu,Sen:1997xi}. For $N=3$ the planarity condition is trivially satisfied at generic points in moduli space. The three D3-branes on which the  string network ends define three points in the $\R^6$ transverse space, which
generically define a plane in this $\R^6$.

It is clear that for $N \geq 4$, the positions of the D3-branes  will not generically be coplanar. If we consider the plane in $\mathbb{R}^6$ defined by the positions of any given three D3-branes, then in order to obtain a planar configuration we need to tune the four transverse positions of each of the remaining D3 branes relative to this plane. Thus a dyonic configuration can  be planar -- and hence supersymmetric -- only on  a constrained submanifold of the moduli space, of codimension $4 (N-3)$.  Even slightly away from this submanifold, the state ceases to be supersymmetric and is thus continuously connected to a nonsupersymmetric state. Hence it must belong to a long multiplet and cannot contribute to the index that counts 64-dimensional intermediate multiplets.

\subsection*{Dyons in String Theory}

To  generalize this argument  to the supergravity situation, it is better to formulate it entirely in terms of the superalgebra. To do so,  consider the low energy effective theory of heterotic string theory compactified on $T^6$. The low energy action consists of four-dimensional $\N=4$ supergravity coupled to 22 vector multiplets. This theory has 134 moduli, which lie in the space
\be
\mathcal{M}=\frac{O(22, 6; \R)}{O(22) \times O(6)} \times \frac{SL(2,\R)}{SO(2)}
\ee
The $\frac{SL(2,\R)}{SO(2)}$ coset is parameterized by the axion-dilaton $\tau= \tau_1 + i \tau_2$, while the $\frac{O(22, 6; \R)}{O(22) \times O(6)}$ coset is parameterized by the remaining 132 moduli. We can encode these moduli in a $28 \times 28$ matrix $M$, satisfying
\be
\CM L \CM^T = L\;, \;\;\;\;\; \CM^T=\CM\;, \;\;\;\;\; \CM \equiv \mu^T \mu \, ,
\ee
where $L$ is the $O(22, 6)$-invariant metric $L= \mbox{diag} ( -I_{22}, I_6)$. The vielbein $\mu$ is identified with $ k \mu$ for any group element $k \in O(22) \times O(6)$, since it defines the same moduli matrix $\CM$. The theory contains 28 gauge fields, with gauge group $U(1)^{28}$ at generic points in the moduli space.

A  dyonic state is specified by its charge vector
\begin{equation}\label{chargevec}
    \Gamma = \left(
               \begin{array}{c}
                 Q \\
                 P \\
               \end{array}
             \right)
\end{equation}
where $Q$ and $P$  are the electric and magnetic charge vectors respectively. Both $Q$ and $P$ are elements of a self-dual integral lattice $\Pi^{22, 6}$ and can be represented as $28$-dimensional column vectors  in $\R^{22, 6}$ with integer entries, which transform in the fundamental representation of $O(22, 6; \Z)$. Given these moduli-independent charge vectors, we can define their moduli-dependent, right-moving projections onto the  spacelike subspace $\R^6$ by
\be
Q_R = \half (I_{28} + L ) \mu^T_{\infty}Q \;, \;\;\;\;\; P_R = \half (I_{28} + L ) \mu^T_{\infty} P \label{defqpr}
\ee
where the subscript ``$\infty$'' refers to the value of the moduli measured at infinity.

In the rest frame of the dyon, the $\N=4$ supersymmetry algebra takes the form
\be
\{Q_{\a}^A, Q_{\dot{\b}}^{\dag B} \} =M  \d_{\a\dot{\b}} \, \d^{AB} \;,\;\;\;\;\; \{Q_{\a}^A, Q_{\b}^{ B} \} = \e_{\a\b} Z^{AB}\;, \;\;\;\;\;\{Q_{\dot{\a}}^{\dag A}, Q_{\dot{\b}}^{\dag B} \} = \e_{\dot{\a}\dot{\b}} \bar{Z}^{AB}
\ee
where $A,B = 1, \ldots 4 $ are $SU(4)$ R-symmetry indices\footnote{We use a convention where the  $A,B$ indices  are raised by complex conjugation. }
 and $\a,\b$ are Weyl spinor indices.
The central charge matrix $Z$ encodes information about the charges and the moduli. To write it explicitly, we first define a central charge vector
in $\C^6$
\be \label{Zvector}
Z^m (\Gamma)  = \frac{1}{\sqrt{\tau_2}}(Q_R^m - \tau P_R^m)   \;, \;\;\;\; m=1,\ldots 6 \;,\;\;\;\;\;
\ee
which transforms in the (complex) vector representation of $Spin(6)$. Using the equivalence $Spin(6) = SU(4)$, we can relate it to the  antisymmetric representation of $Z_{AB}$ by
\begin{equation}\label{defZ}
Z_{AB}(\Gamma) = \frac{1}{\sqrt{\tau_2}} (Q_R - \tau P_R)^m \lambda^{m}_{AB} \;, \;\;\;\;\; m=1, \ldots 6
\end{equation}
where $\lambda^{m}_{AB}$ are the Clebsch-Gordan matrices.
An explicit representation for $\lambda^m_{AB}$ is given in the appendix.
Since $Z(\Gamma)$ is antisymmetric, it can be brought to a block-diagonal form  by a $U(4)$ rotation\begin{equation}\label{diag}
\widetilde{Z} = U Z U^T, \;\; U \in U(4)\;, \;\;\;\;\; \widetilde{Z}_{AB} = \left(\begin{array}{c|c} Z_1 \varepsilon & 0  \\   \hline  0 & Z_2 \varepsilon \end{array}\right)\;, \;\;\; \varepsilon = \left(\begin{array}{cc} 0 & 1  \\  -1 & 0 \end{array}\right)
\end{equation}
where $Z_{1}$ and  $Z_2$ are non-negative real numbers. This rotation acts on the supercharges as
\be\label{qrot}
\widetilde{Q}^A = Q^B (U^\dagger)_B^A \, .
\ee
Since $\varepsilon$ is the invariant tensor of $SU(2)$, a $U(2) \times U(2)$ transformation acting separately on each block can only change the phases of $Z_1$ and $Z_2$. We will therefore be more general and treat $Z_1$ and $Z_2$ as complex numbers. Without loss of generality we can assume $|Z_1| \ge |Z_2|$.

We now split the $SU(4)$ index as $A=(a,i)$, where $a,i=1,2$ and $i$ represents the block number.
Defining the following fermionic oscillators
\begin{equation}\label{defAB}
    \mathcal{A}_{\a}^{ i} = \frac{1}{\sqrt{2}} ( \widetilde{Q}^{1i}_\alpha+ \epsilon_{\a \b} \widetilde{Q}^{\dag \, 2 i }_\b ), \quad \widetilde{B}_{\a}^{ i} = \frac{1}{\sqrt{2}} ( \widetilde{Q}^{1i}_\alpha - \epsilon_{\a \b} \widetilde{Q}^{ \dag \, 2 i}_\b ) \;,
\end{equation}
the supersymmetry algebra takes the form
\be\label{ABalg}
\{\mathcal{A}_{\dot{\a}}^{ i \dag} , \mathcal{A}_{\b}^{ j}\} = (M+ Z_i) \,\d_{\dot{\a}\b} \,\d^{ij } \;, \;\;\;\;\;\{\mathcal{B}_{\dot{\a}}^{ i \dag} , \mathcal{B}_{\b}^{ j}\} = (M- Z_i) \,\d_{\dot{\a}\b}\, \d^{ij}
\ee
with all other anti-commutators being zero.

If $M > |Z_1| > |Z_2|$, no supersymmetries are preserved.
The sixteen broken supersymmetries result in eight complex fermionic zero modes  whose quantization furnishes a $2^{8}$-dimensional  long multiplet. If  $M = |Z_1| > |Z_2|$, the state is quarter-BPS,  and four out of the sixteen supersymmetries are preserved. The broken supersymmetries result in six complex fermionic zero modes whose quantization furnishes a $2^{6}$-dimensional intermediate multiplet.  If  $M = |Z_1| = |Z_2|$, the state is half-BPS,  and eight out of the sixteen supersymmetries are preserved. The broken supersymmetries result in four complex fermionic zero modes whose quantization furnishes a $2^{4}$-dimensional short multiplet.

The supersymmetries preserved by a given state is thus specified by the central charge matrix.
Furthermore, given a quarter BPS state, its charges pick out a  particular $\N=2$ subalgebra of the $\N=4$ algebra -- in the above basis, this subalgebra is generated by
$(\mathcal{A}^{ 1},\mathcal{B}^{ 1})$ and their complex conjugates.
A quarter-BPS configuration of the $\CN=4$ algebra is a half-BPS configuration of this $\CN=2$ subalgebra, and it is annihilated  by the supersymmetry generated by\footnote{The supersymmetry transformation on the fields with variation parameter $\epsilon_{\a}$ induced by a supercharge $Q_{\a}$ is $\epsilon^{\alpha} Q_{\alpha} + \epsilon^{\dagger \alpha} Q^{\dagger}_{\alpha}$.}
$\mathcal{B}^{ 1}$.

It will be useful later to state all this in a more covariant form. In the above basis, the preserved supercharge, after allowing for a $U(1)$ rotation in the $\CN=2$ subalgebra, can be written as\footnote{Here we have suppressed spacetime spinor indices; each element of $Q$ is a spacetime spinor.}:
\be\label{Qtlcan}
\widetilde{Q}_A= \left(  e^{i \theta},  -\gamma_0 e^{-i \theta}, 0, 0 \right) \, ,
\ee
and it obeys  the projection equation:
\be\label{projQtl}
\gamma_0 \widetilde{Q}_A =  \widetilde{Q}^A \left( \begin{array}{c|c} \varepsilon^{T} & 0 \\ \hline 0 & 0 \end{array} \right) \, .
\ee
To write this projection condition in a  covariant form, we transform back to the original basis:
\be\label{Zproj}
\gamma_0 Q_A =   Q^B \frac{\hat{Z}^{T}_{AB}}{|\hat{Z}|} \, ,
\ee
where
\be\label{defZhat}
\hat{Z}_{AB}
 =  U^{\dagger} \left( \begin{array}{c|c} Z_1 \varepsilon & 0 \\ \hline 0 & 0 \end{array} \right)
 U^{*}
\ee
and
\be\label{defmodZ}
|\hat{Z}| = (Tr[\hat{Z} (\hat{Z})^{\dag}])^{\half} = Z_1.
\ee

It is useful to think of the central charge as a vector $Z^m$  instead of as an antisymmetric matrix $Z_{AB}$. Given an  orthonormal basis $\{e_A\}$  for the $\bf 4$ of $SU(4)$, the basis for the $\bf 6$ is given by the tensors $\{ e_A \wedge e_B\}$, $A, B= 1,\ldots, 4$. This is related to the  orthonormal basis $\{ f^m \}, \, m = 1, \ldots 6$ for the vector representation of $Spin(6)$ by
\begin{equation}\label{basischange}
    f^m =  (e_A \wedge e_B) \, \lambda^m_{AB}\, .
\end{equation}
Now, an  $SU(4)$ rotation which rotates the supercharges, $Q'=Q U $, acts on the Clebsch-Gordan matrices as
\be\label{transform}
U \lambda^m U^T = R^m{}_n(U) \lambda^m
\ee
where $R^m{}_n$ is an $SO(6)$ rotation matrix. In particular, the diagonalization \eqref{diag} simply rotates the central charge vector $\vec{Z}$ to the $(12)$ plane, where it takes the form
\be
\vec{\tilde{Z}} = (Z_1+Z_2) {f}_1 + i(Z_1-Z_2) {f}_2 \, . \label{canon}
\ee
Since $Z^m$ is a (complex) linear combination of the charge vectors $Q_R$ and  $P_R$ given by (\ref{Zvector}), this transformation simply rotates the plane spanned by the charge vectors $Q_R$ and $P_R$ into  the plane spanned by  the basis vectors $f_1$ and $f_2$.

To recap, the $\CN=2$ subalgebra relevant for discussing the BPS-properties of a given state is  completely determined by the plane spanned by the charge vectors\footnote{Note that for a half-BPS state ${Q}_R$ and ${P}_R$ are proportional, and the plane degenerates to a line.} $Q_R$ and $P_R$.
Given the preserved supercharge of the solution, one finds the matrix $U$ which rotates it into the canonical form \eqref{Qtlcan}. In $\IR^{6}$, the same transformation rotates the plane spanned by $(f_{1},f_{2})$ into the plane spanned by $(Q_{R}, P_{R})$ in $\IR^{6}$.
The matrix $U$ is not unique, but given two matrices $U_{1}, U_{2}$ which rotate the supercharges satisfying the condition \eqref{Zproj} into the canonical form \eqref{Qtlcan}, the
matrix $\Phi =  U_1 (U_2)^{-1} $ is unitary and satisfies
\be\label{Phiprop}
\widetilde{Q}^{A} \Phi  = \widetilde{Q}^{A}
\ee
where $\widetilde{Q}_A$ is the canonical form \eqref{Qtlcan}.
This condition implies that $\Phi$ is block-diagonal, with a $U(2)$ matrix in each block, and the action on $Z$ reduces to two independent $U(1)$ actions rotating the  phases of $Z_{1}$ and $Z_{2}$.
The plane determined in $\IR^{6}$ is thus unambiguous.

\bigskip

Now consider a two-centered solution of the $\N=2$ theory where both centers are large. Let the charges of the centers be $(Q^{1}, P^{1})$  and  $(Q^{2}, P^{2})$. If this configuration is embedded supersymmetrically in the $\CN=4$ theory by truncating the latter theory, it is clear that the two centers have to pick the same $\CN=2$ algebra inside $\CN=4$. By the above argument, the planes defined by $(Q_{R}^{1}, P_{R}^{1})$  and  $(Q_{R}^{2}, P_{R}^{2})$ must coincide. This happens only on a submanifold of moduli space, as in the field theory analysis.

As mentioned above, there could exist solutions in the $\CN=4$ theory that cannot be truncated to $\CN=2$ theory. To extend the above argument to these more general multi-centered configurations,  we will make an approximation that one of the black holes has small charges and can be considered to be a probe in the background of the other big black hole\footnote{In order to rule out the unlikely possibility that supersymmetry is broken in the probe approximation but is regained after backreaction of the second center, one must analyze the full $\CN=4$ Killing spinor equations in the multi-center system. We shall not do this in the present paper. We shall comment on this issue later.}.

\subsection*{A dyonic probe in the background of a dyonic black hole}

The spirit of the argument is again that the two centers generically break different sets of supersymmetries. However, since there is no longer any special globally defined $\CN=2$ subalgebra, we have to rephrase our arguments using local supersymmetry, {\it i.e.} Killing spinors, which we shall denote by $\e^A (x)$.

Consider then a single centered, quarter-BPS dyonic black hole solution with (super)charges as described above.
We define a {\it local} version  $Z^b_{AB}(x)$ of the central charge matrix \eqref{defZ} using the charges of the black hole and the local values of the moduli fields. One can block-diagonalize this matrix as in \eqref{diag} using a local matrix $U_{b}(x)$. In this way, we obtain $\widetilde Z^{b} (x) \equiv U_{b} (x) Z^{b} (x) U_{b}^{T} (x)$ and the two complex numbers $(Z_1^b, Z_2^b)$ with $|Z_{1}^{b}| > |Z_{2}^{b}|$.
It turns out \cite{Billo:1999ip} that the Killing spinor of the background satisfies a projection condition at each point in spacetime:
\be\label{Zbproj}
\gamma_0 \e_A(x) = \frac{\hat{Z}^b_{AB}(x)}{|\hat{Z}^{b}(x)|} \e^B(x) \;, \;\;\;\;\; \e^A = (\e_A)^{\star} \, ,
\ee
where
$\hat{Z}_{AB}$ and $|\hat{Z}^{b}|$ are defined as in \eqref{defZhat}, \eqref{defmodZ} respectively, but using the matrix $U_{b}(x)$ instead of $U$.

A probe placed  at some point $x_0$ in this background will generically break the existing supersymmetries. We define the local quantities $Z^{p}(x), U_{p}(x), \widetilde{Z}^{p}(x), \hat{Z}^{p}(x)$ as above, but using the charges of the probe  everywhere.
A $\kappa$-symmetry analysis on the probe worldvolume \cite{Billo:1999ip} shows that it will preserve the background supersymmetries only if the following condition is met\footnote{In \cite{Billo:1999ip} the authors were dealing with $\CN=8$ supergravity  backgrounds, but here we assume that their expressions particularize straightforwardly to the $\CN=4$ case. }
\be\label{ksymm}
\gamma_0 \e_A(x_0) = \frac{\hat{Z}^p_{AB}(x_0)}{|\hat{Z}^{p}(x_0)|} \e^B(x_0) \, .
\ee

We thus have the two projection conditions \eqref{Zbproj},\eqref{ksymm} on the Killing spinor $\e_A(x_{0})$ at the location of the probe. These are local analogs of the projection equation\footnote{Strictly speaking, since $\e^{A} Q_{A}$ is a scalar, there is a transposition involved in these equation with respect to \eqref{Zproj}.} \eqref{Zproj} and involve the two matrices $U_{p}(x_{0})$ and $U_{b}(x_{0})$.
As argued above, this implies that the plane in $\IR^{6}$ spanned by the charges $(Q^{p}_{R}, P^{p}_{R})$ of the probe must be the same as the plane spanned by the charges $(Q^{b}_{R}, P^{b}_{R})$ of the background.
We can now determine the BPS properties of a multi-centered configuration geometrically, directly in the charge space instead of in the space of supercharges.

Recall that for $\CN=2$ multicentered bound states, there is one supersymmetry requirement  (the alignment of the phases of the central charges of the two centres), and one tunable parameter (the radial distance), which is a first step in showing that multicentered solutions in $\CN=2$ exist generically in the vector multiplet moduli space  \cite{Denef:2000nb}.
In the $\CN=4$ theory,  there  are three possibilities for two-centered solutions:
\begin{myenumerate}
\item Both centers are half-BPS. In this case, the plane defined by each center degenerates into a line. Since two lines  are then trivially coplanar, it is always possible  to identify an $\CN=2$ subalgebra without adjusting any parameters. What remains is to align the preserved $\CN=1$ supersymmetry (the one remaining phase) and there is one tunable parameter $x_{0}$, the distance between the centers, which one can use to do so. This is analogous to the case of $SU(3)$ dyons in field theory.

\item One center is half-BPS and the other is quarter-BPS. In this case, one needs to align  the line corresponding to the half-BPS center with
the plane corresponding to the quarter-BPS center. This will only happen in a constrained locus in the moduli space, obtained by setting  to zero the four components of the line that are perpendicular to the plane.  This is analogous to the case of $SU(4)$ dyons in field theory.

\item Both centers are quarter-BPS. The two planes corresponding to the two centers will coincide on a locus in the moduli space  even more constrained than in the previous case and will require tuning eight  parameters. This is analogous to the $N >4$ cases for  $SU(N)$ dyons in field theory\footnote{We need to tune only eight parameters and not $4(N-3)$  because we are considering the situation when each center is already independently quarter-BPS.}.
\end{myenumerate}
We therefore conclude that the multi-centered configurations of the second and the third type, which contain at least one quarter-BPS center, exist as supersymmetric configurations only on submanifolds of  codimension larger than one. This implies that  they are smoothly connected to nonsupersymmetric long multiplets.

\medskip

While we have derived  the constraints arising from supersymmetry alignment
 only in the probe approximaton, it is unlikely that including backreaction would restore supersymmetry away from the constrained locus.
In order to argue this rigorously, one must analyze the $\CN=4$ supersymmetry equations. One may also be able to devise a much simpler argument based on requiring that there exist\footnote{In the $\CN=2$ situtation, there is indeed always such a point in spacetime \cite{Denef:2000nb}.} a point in spacetime where the Killing spinor  simultaneously obeys two projection equations  similar to \eqref{Zbproj}, \eqref{ksymm}. Our analysis of supersymmetry alignment will then go through as above,  with the only change that $x_{0}$ is no longer identified as the location of the probe.

\subsection*{Discussion}

Our analysis is independent of but consistent with the analysis of walls of marginal stability \cite{Sen:2007vb, Dabholkar:2007vk,Cheng:2007ch} and the analysis of rare decay modes \cite{Sen:2007nz,Mukhi:2008ry}. The states of the first type do contribute to the dyon partition function. These two-centered solutions decay on a wall  of marginal stability of codimension one, where the distance between the two centers goes to infinity. The dyon partition function has poles in precise correspondence with these walls of marginal stability. On the other hand, the states of the second and third type contribute zero  to the dyon partition function everywhere in the moduli space. This is consistent with the fact that the dyon partition function has no additional singularities other than the poles described above and there are no additional jumps in the degeneracies. If the dyon partition function did count these states, it would have to display an unusual singularity structure, since these states decay on surfaces of codimension two or higher.

Our  arguments could also be made in $\CN=4$ theories in five dimensions to rule out the contribution of enigmatic configurations to the indexed degeneracy. This is consistent with the fact that the entropy of the single-centered three-charge black holes in type IIB string theory on $K3 \times S^{1}$  agrees to sub-leading order with the microscopic indexed degeneracy in the {\it same} regime of charges that are relevant for black hole \cite{Castro:2008ys}.

Finally, our analysis raises the following question: if the usual index does not count the two-centered configurations with at least one  quarter-BPS center, can one devise some other method to count them on the submanifold where they do exist as supersymmetric configurations? One possible method is to use an argument employed in \cite{Dabholkar:2008tm}. The degeneracy of the bound state can be computed by first going very close to the line of marginal decay, where the two centers are very far away. In this regime, it is simply the product of the degeneracies of each center, times a multiplicity factor coming from the field angular momentum. The degeneracy of each quarter-BPS center is in turn computed using the usual dyon partition function for quarter-BPS states. By continuity of the index, the degeneracy of the dyonic state does not change if we move far away from the line of marginal decay.

\subsection*{Acknowledgments}

We would like to thank Micha Berkooz, Iosif Bena, Gabriel Cardoso, Alejandra Castro, Miranda C.\,N.\,Cheng, Roberto Emparan, Ilies Messamah, Boris Pioline, and Ashoke Sen for useful discussions. S.N. would like to thank the CNRS and the string theory group at the LPTHE for hospitality.
The work of A.D. is supported in part by the Excellence Chair of the Agence Nationale de la Recherche (ANR). S.M. is supported by the European Commision Marie Curie Fellowship under the contract PIIF-GA-2008-220899.
S.N. is supported in part by the European Community's Human potential
program under contract MRTN-CT-2004-005104 ``Constituents, fundamental
forces and symmetries of the universe'', the Excellence Cluster ``The
Origin and the Structure of the Universe'' in Munich and the German
Research Foundation (DFG) within the Emmy-Noether-Program (grant
number: HA 3448/3-1).

\subsection*{Appendix}
The Clebsch-Gordan matrices $\lambda^m_{AB}$ are given by the components $(C\Gamma^m)_{AB}$ where $\Gamma^m$ are the Dirac matrices of $Spin(6)$ in the Weyl basis satisfying the Clifford algebra $\{ \Gamma^m , \Gamma^n \} = 2 \delta^{mn}$, and $C$ is the charge conjugation matrix.  The Gamma matrices are given explicitly in terms of Pauli matrices by
\begin{eqnarray}
  \Gamma^1 = \sigma_1 \times \sigma_1 \times 1 &, \quad& \Gamma^4 = \sigma_2 \times 1 \times \sigma_1 \, , \\
  \Gamma^2 = \sigma_1 \times \sigma_2 \times 1&, \quad& \Gamma^5 =  \sigma_2 \times 1 \times \sigma_2  \, , \\
  \Gamma^3 = \sigma_1 \times \sigma_3 \times 1&, \quad& \Gamma^6 = \sigma_2 \times 1 \times \sigma_3 \, .
\end{eqnarray}
The charge conjugation matrix is defined by
$C\Gamma^m C^{-1} = - {\Gamma^m}^*$
\begin{equation}\label{chargeconj}
    C = \sigma_1 \times \sigma_2 \times\sigma_2,
    \quad C\Gamma^m = \left(
                                           \begin{array}{cc}
                                             \lambda^m_{AB} & 0 \\
                                             0 & {\bar\lambda}^{m}_{{\dot A} {\dot B}} \\
                                           \end{array}
                                         \right)
\end{equation}
where the un-dotted indices transform in the spinor representation of $Spin(6)$ or the $\bf 4$ of $SU(4)$ whereas the the dotted indices transform in the conjugate spinor representation of $Spin(6)$ or the $\bf \bar 4$ of $SU(4)$. The matrices $\lambda^m _{AB}$ thus defined have the required antisymmetry and  transform properties as in (\ref{transform}).

\bibliographystyle{utphys}
\bibliography{nonenigma}

\end{document}